\newcommand{\tindex}[1]{{\scriptstyle {\rm#1}}}

\documentclass{mystyle}
\usepackage[xdvi]{graphicx}
\usepackage{amssymb}
\usepackage[]{epic,eepic,epsf}
\begin{document}
\begin{frontmatter}

  \title{Thermally activated interface motion in a disordered ferromagnet}

  \author{L.\ Roters},
  \corauth[cor]{Corresponding author}
  \ead{lars@thp.Uni-Duisburg.DE}
  \author{S.\ L\"ubeck} and
  \ead{sven@thp.Uni-Duisburg.DE}
  \author{K.\,D. Usadel}
  \ead{usadel@thp.Uni-Duisburg.DE}
  \address{
    Theoretische Physik,
    Gerhard-Mercator-Universit\"at,
    Lotharstr. 1,
    47048 Duisburg,
    Germany
    }
\begin{abstract}
  We investigate interface motion in disordered ferromagnets by
  means of Monte Carlo simulations. For small temperatures and driving
  fields a so-called creep regime is found and the interface velocity
  obeys an Arrhenius law. We analyze the corresponding energy barrier
  as well as the field and temperature dependence of the prefactor.
\end{abstract}
\begin{keyword}
  disordered media \sep
  driven interfaces \sep 
  creep motion 
  
  \PACS 68.35.Rh \sep 68.35.Ja \sep 75.10.Hk \sep 75.40.Mg
\end{keyword}
\end{frontmatter}
\section{Introduction}
\label{intro}
Using the random-field Ising model (RFIM) we study the dynamics of
driven interfaces in disordered ferromagnetic systems by means of
Monte Carlo simulations.
The interface separates regions of opposite magnetization and it is
driven by a homogeneous field $H$.
Without thermal fluctuations ($T=0$) its dynamics is affected by the
disorder in the way that the interface moves only if a sufficiently
large driving field is applied.
With decreasing driving field the interface velocity vanishes
at a critical threshold $H_\tindex{c}$ and it is pinned below
$H_\tindex{c}$, due to the random-field. 
The field dependence of the interface velocity is given by 
\begin{equation}
  v ( h ) \sim h^\beta
  \label{eq:d3}
\end{equation}
with $h=(H-H_\tindex{c})/H_\tindex{c} \geq0$, 
i.e.\ the pinning/depinning transition can be considered as a
continuous phase transition.

For finite temperatures $T>0$ no pinning of the interface occurs
since the energy needed to overcome local pinning centers can be
provided by thermal fluctuations. 
In particular for driving fields well below $H_\tindex{c}$ the
interface velocity remains finite and is expected to exhibit a
so-called creep motion in which the interface velocity obeys an
Arrhenius law. 
This Arrhenius law has been found for the Edwards-Wilkinson
equation with quenched disorder~\cite{CHAUVE_1,CHAUVE_2} 
and we compare these results with those exhibited by a driven
interface in the 3$d$~RFIM.
The Hamiltonian of this model is given by
\begin{equation}
  {\bf H} =
  -\frac{J}{2}\, \sum_{\langle i,j \rangle} S_i \, S_j
  -H \, \sum_{i} S_i
  - \sum_{i} h_i\,S_i
  \mbox{ .}
\end{equation}
The sum in the first term is taken over all pairs of neighbored spins.
It describes the exchange interaction between neighboring
spins ($S_i=\pm 1$) and a parallel alignment of nearest neighbors is
energetically favored. 
Additionally, the spins are coupled to a homogeneous driving field $H$
and to a quenched local random-field $h_i$ with
$\langle h_i h_j \rangle \propto \delta_{ij}$
and $\langle h_i \rangle = 0$.
The probability density $p$ that the
local random-field takes some value $h_i$ is given by
$p(h_i) = (2\Delta)^{-1}$ for $|h_i| < \Delta$ and zero otherwise. 
Using simple cubic lattices with antiperiodic boundary conditions
we started each simulation with an initially
flat interface. 
In non-disordered systems the interface moves for any
finite driving field. 
This limiting behavior can be recovered if the interface moves along
the diagonal direction of simple cubic lattices (see for
details~\cite{NOWAK_1}).  
In our Monte Carlo simulations the 
dynamics of the interface motion is given by a Glauber
dynamics with transition probabilities according to a heat-bath
algorithm (see e.g.~\cite{BINDER_1}). 
The basic quantity in our investigations is the interface velocity. 
Since a moving interface corresponds to a magnetization $M$ which
increases with time $t$ (given in Monte Carlo steps per spin) 
the interface velocity is obtained 
according to $v \sim \langle {\rm d}M/{\rm d}t \rangle$ where
$\langle ... \rangle$ denotes an appropriate disorder average. 
\section{Creep motion}
\label{creep}

The depinning transition occurs at zero temperature only. 
Including thermal fluctuations no pinning of the interface takes
place. 
At the critical threshold itself the velocity grows with the
temperature according to
$v(H=H_\tindex{c}) \sim T^{1/\delta}$~\cite{NOWAK_1,ROTERS_1}.
Well below $H_\tindex{c}$ the interface velocity is expected
to obey an Arrhenius 
law 
\begin{equation}
  v \sim C(H,T) \, \exp \left[-\frac{E(H)}{T}\right]
  \label{eq:arr}
\end{equation}
which is characterized by its prefactor $C(H,T)$ and an effective
energy barrier~$E(H)$. 
Recent investigations of the creep regime in the QEW yield that the
prefactor of the Arrhenius law is given by~\cite{CHAUVE_1}
\begin{equation}
  C(H,T)=c(H) \, T^{-x}
  \label{eq:pre}
\end{equation}
with $c(H)$ being some function of the driving field and the
temperature dependence is characterized by an exponent $x$.
We apply this ansatz to our numerically obtained interface velocities
in the 3$d$ RFIM. 
\begin{figure}[t]
 \begin{center}
   \includegraphics[height=7cm]{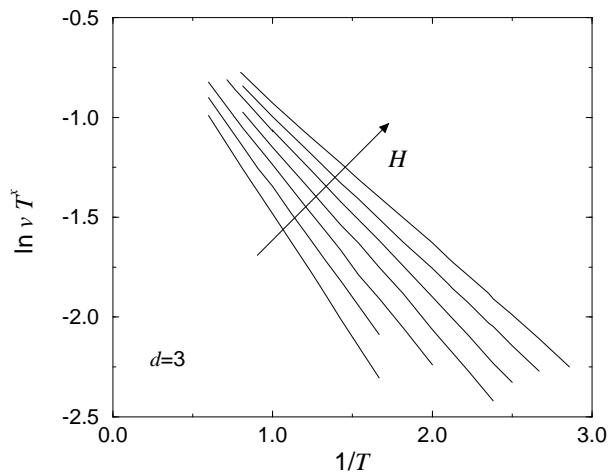} 
   \caption{
      {\footnotesize
     Interface velocities as a function of the inverse temperature for
     different values of the driving field
     $H =0.3, 0.35, 0.4, ..., 0.6$.
     The velocities are rescaled according to
     Eqs.~(\ref{eq:arr},\ref{eq:pre}).
     Varying $x$ we obtain nearly straight lines for
     $x=0.79 \pm 0.09$. 
     }}
   \label{fig:lines}
 \end{center}
\end{figure}
Figure~\ref{fig:lines} shows the data rescaled according to 
Eqs.~(\ref{eq:arr},\ref{eq:pre}).
As can be seen from the data, we obtain nearly straight lines for
$x=0.79\pm 0.09$ suggesting that the temperature dependence of the
prefactor as well as the Arrhenius law itself are a proper
description of the velocity in the creep regime.  
The slope of the lines in Fig.~\ref{fig:lines} corresponds to the
effective energy barrier $E(H)$.
Plotting $\ln v\,t^x$~vs.~$E(H)/T$ the different curves collapse onto
a single curve (see Fig.~\ref{fig:coal}) without making any assumption
about $c(H)$. 
This suggests that $c(H) \approx {\rm const}$. 
\begin{figure}[t]
 \begin{center}
   \includegraphics[height=7cm]{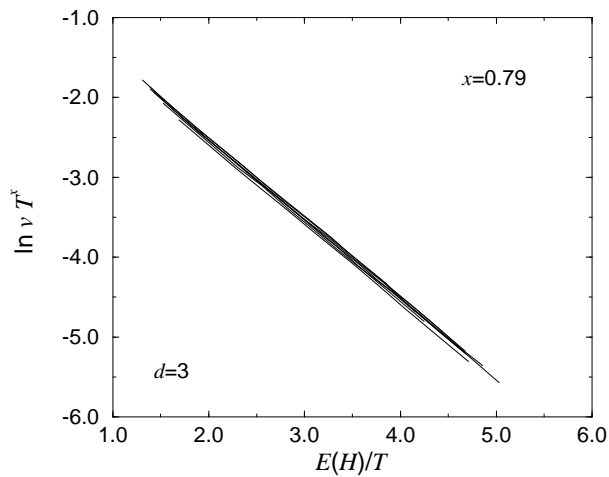} 
   \caption{
      {\footnotesize
     Rescaled interface velocities.
     The data coincide with those shown in Fig.~\ref{fig:lines}.
     They are rescaled with the numerically determined energy
     barrier~$E(H)$.
     The coalescence of the data occurs without any consideration of
     $c(H)$. This indicates that $c(H) \approx {\rm const}$. 
     }}
   \label{fig:coal}
 \end{center}
\end{figure}

\section{Summary}

We study the creep regime exhibited by a driven interface in the
random-field Ising model. 
The creep regime occurs for driving fields well below the critical
threshold and small but finite temperatures. 
We find that the interface velocity obeys an Arrhenius law. 
Inspired by a renormalization group approach~\cite{CHAUVE_1} we
investigate the details of this Arrhenius law.
Studying the corresponding prefactor [Eq.~(\ref{eq:pre})] we find
that its details do not coincide with~\cite{CHAUVE_1}. 
In particular, these authors observed a significant field dependence of
the corresponding prefactor $c(H) \sim H^\mu$ with $\mu>0$ and obtained a
negative exponent $x$. 
Especially the opposite sign of $x$ is remarkable and
further investigations are needed to clarify this point. 
\section*{Acknowledgments}
This work was supported by the Deutsche Forschungsgemeinschaft via
Graduiertenkolleg {\it Struktur und Dynamik heterogener Systeme}
at the University of Duisburg.

\end{document}